\documentclass[draft]{agujournal2019}
\usepackage{url} 
\usepackage{lineno}
\usepackage{amsmath}
\usepackage[inline]{trackchanges} 
\usepackage{etoolbox}
\let\bbordermatrix\bordermatrix
\patchcmd{\bbordermatrix}{8.75}{4.75}{}{}
\patchcmd{\bbordermatrix}{\left(}{\left[}{}{}
\patchcmd{\bbordermatrix}{\right)}{\right]}{}{}

\usepackage{soul}
\DeclareMathOperator*{\argmin}{\arg\!\min}

\draftfalse

\journalname{Geophysical Research Letters}

\begin{document}

\title{Through a Jet Speed Darkly: The Emergence of Robust Euro-Atlantic Regimes in the Absence of Jet Speed Variability}

%
%

\authors{J. Dorrington\affil{1} and K. J. Strommen\affil{1}}

\affiliation{1}{Department of Physics, University of Oxford, Oxford, UK}

\correspondingauthor{Kristian J. Strommen}{kristian.strommen@physics.ox.ac.uk}

\begin{keypoints}
\item The approximately Gaussian jet speed serves to obscure non-Gaussian regime structure in the phase space of geopotential height.
\item If this influence is removed, the regime structure becomes significantly more robust and stable across the entire 20th century.
\item We find a new paradigm of 3 main regimes that can be consistently extended to 5 regimes, capturing both jet latitude and blocking patterns.
\end{keypoints}


\begin{abstract}
Euro-Atlantic regimes are typically identified using either the latitude of the eddy-driven jet, or clustering algorithms in the phase space of 500hPa geopotential height (Z500). However, while robust trimodality is visibly apparent in jet latitude indices, Z500 clusters require highly sensitive significance tests to distinguish them from autocorrelated noise. As a result, even small shifts in the time-period considered can notably alter the diagnosed regimes. Fixing the optimal regime number is also hard to justify. We argue that the jet speed, a near-Gaussian distribution projecting strongly onto the Z500 field, is the source of this lack of robustness. Once its influence is removed, the Z500 phase space becomes visibly non-Gaussian, and clustering algorithms easily recover three extremely stable regimes, corresponding to the jet latitude regimes. Further analysis supports the existence of two additional regimes, corresponding to a tilted and split jet. This framework therefore naturally unifies the two regime perspectives.
\end{abstract}

\section*{Plain Language Summary}

Weather over the North Atlantic region during Winter is highly variable, with shifts in the jet stream and anticyclonic blocks both having large impacts downstream on Western Europe. A common way of thinking about this variability is in terms of movement between persistent large-scale weather patterns termed \emph{regimes}. However settling on an optimal and consistent set of regime patterns has proved very challenging. We show that by removing the confounding influence of jet speed from the geopotential height field, it is much easier to identify stable regime patterns which well describe observations. These patterns include both jet dynamics and blocked states, forming a bridge between the sets of previous studies that looked for regimes in either the jet latitude or geopotential height in isolation.

%
%
\section{Introduction}
\label{sec:intro}

One way to understand non-linear variability in the Euro-Atlantic circulation is through the study of non-Gaussian structure in its phase space, which indicates preferred flow configurations. Since the 1980's, several studies have found evidence to suggest the existence of such deviations from Gaussianity, manifesting themselves in the form of quasi-persistent weather regimes in the Euro-Atlantic region (e.g. \citeA{Vautard1990} and \citeA{Michelangeli1995} and discussion within for early history; more recently see e.g. \citeA{Straus2007}, \citeA{Cassou2008}, \citeA{Straus2010}, \citeA{Woollings2010a}, \citeA{Woollings2010b}, \citeA{Franzke2011} and \citeA{Hannachi2017}). Their importance in modulating European weather is now well documented \cite{Frame2013, Ferranti2015, Matsueda2018} and, conjecturally, they may even critically influence the regional response to anthropogenic forcing \cite{Palmer1999, Corti1999}. From the perspective of weather forecasting, they offer a potentially vastly simplified truncation of the atmosphere to Markovian dynamics \cite{Ghil2002d, Strommen2019}.

There are two main approaches in the literature for diagnosing Euro-Atlantic regimes. The most commonly used method is to apply clustering algorithms to geopotential height data at 500 hPa (Z500). This typically produces 4 regimes: the positive and negative North Atlantic Oscillation (NAO), an Atlantic ridge and a Scandinavian blocking pattern  \cite{Cassou2008, Dawson2012, Strommen2019a}. However, two problematic issues emerge in this framework. Firstly, the choice of 4 regimes in particular depends on complicated statistical significance testing, which is typically highly sensitive to the inclusion or removal of small numbers of points. For reanalysis data, a shift in the time-period considered by as little as 5 years can mean the difference between significance or not \cite{Strommen2019a}. For models, which appear to struggle to replicate good regime structure, this sensitivity is considerably magnified (ibid). Secondly, as we will show in this paper, the 4 regimes are highly unstable across the 20th century, exhibiting significant decadal fluctuations in their spatial patterns. While this may ostensibly be a feature of the climate system, it calls into question the use of these regimes for any practical purposes.

Arguably, the main underlying issue responsible for all this ambiguity is the fact that the Z500 phase space is not easily visually distinguishable from Gaussian noise (see e.g. Figure 1a of this paper). This is in stark contrast to the other common way of identifying Euro-Atlantic regimes by appealing to the eddy-driven jet. Indeed, in \citeA{Woollings2010}, a simple computation of the daily latitude of the jet produces a visibly trimodal histogram, suggesting the existence of 3 distinct jet regimes without the need for further significance testing. Furthermore, these jet regimes are notably stable, showing little sensitivity to the choice of time-period used or subsetting of data. The work in \citeA{Madonna2017} went some way towards reconciling these with the Z500 regimes discussed above, by identifying the 3 jet regimes with 3 of the 4 Z500 regimes. It was argued that the `missing' 4th regime (Scandinavian blocking) corresponds to a tilted jet, not captured by the strictly zonal jet latitude index. However, an obvious question remains: if these two perspectives are capturing the same underlying structure, why are stable and robust Z500 regimes so much harder to diagnose than the corresponding jet regimes?

In this paper we argue that the confounding factor obscuring the non-Gaussian structure in the phase space of Euro-Atlantic Z500 is the jet speed. It is well known that the NAO, the leading mode of variability (i.e. first principal component) of Z500 during DJF, enjoys strong linear correlation with the jet speed, which is, to good approximation, Gaussian in its distribution \cite{Parker2019}. In fact, this is the case for the other principal components as well, and once the influence of the jet speed has been regressed out, we show that one is left with a visibly non-Gaussian phase space. This non-Gaussianity corresponds precisely to the non-linear influence of the jet latitude on the NAO and East Atlantic pattern. By applying K-means clustering algorithms to this new phase space, we find clear support for either 3 or 5 regimes. The choice K=3 corresponds naturally to the three jet latitude regimes, while K=5 adds regimes corresponding to a split and tilted jet, thereby completing the bridge between the jet and Z500 regimes. In both cases, the patterns are remarkably stable across the entire 20th century.

We argue that removing the influence of the jet speed is therefore a natural and highly beneficial step when analysing Euro-Atlantic circulation, and adds considerable additional clarity to how the two regime pictures are related.

In Section \ref{sec:data_and_methods} we present the data used and describe the relevant computational methodologies. In Section \ref{sec:results} we present the results and in Section \ref{sec:conclusions} draw our conclusions.

\section{Data and Methods}
\label{sec:data_and_methods}
\subsection{Data}

Because we will be interested in understanding the stability of regimes on decadal timescales, the long reanalysis dataset ERA20C \cite{Poli2013} is used, which covers the period 1900-2010. ERA20C assimilates only surface observations, but because both Z500 and the jet are tightly constrained by surface pressure alone, its representation of both is not appreciably different from reanalysis data assimilating satellite data. Indeed, using the full time period reproduces the same regime behavior, both in terms of the jet and Z500, more commonly identified using modern reanalysis datasets covering the period 1979 onwards. This dataset, also used in \citeA{Parker2019}, was therefore deemed suitable for our purposes.

We will be working with daily 500 hPa geopotential height fields (Z500) and 850 hPa zonal wind fields (U850) covering boreal winter (DJF) 1901/1902 to 2010/2011.

\subsection{Methods}
\label{sec:methods}

A low-dimensional truncation of the full Z500 phase space is obtained as in \citeA{Dawson2012}. Concretely, we calculate the leading 10 principal components (PCs) of area-weighted Z500 restricted to the Euro-Atlantic sector [80W-40E, 30N-90N], with a seasonal cycle removed from each gridpoint. The computed PCs explain 83.5 \% of the total variance. To make plots more readable, the PCs were standardised by dividing all of them by $10^8$.

Jet latitude and jet speed timeseries are computed using the simplified methodology described in \citeA{Parker2019}. Daily zonal means of zonal winds are restricted to the region [0-60W, 15N-75N] and 850hPa, before being smoothed with a 9-day running mean. For each day of the DJF season, the speed of the jet is defined as the maximum wind speed attained in this domain and the latitude of the jet is the latitude at which this maximum is located. As noted in ibid, this procedure produces qualitatively similar results to more complex methods using winds at multiple levels and/or additional filtering.

Clustering is always done with a standard K-means algorithm, which produces clusters that maximize the \emph{optimal ratio}:
\begin{equation}
\frac{\textit{Inter-cluster variance}}{\textit{Intra-cluster variance}},
\end{equation}
where the inter-cluster variance refers to the variance between the cluster centroids (weighted by the number of points in each cluster), and the intra-cluster variance refers to the average variance of the differences between the cluster centroids and the data-points associated to that cluster. A large inter-cluster variance therefore implies that the centroids are well separated from each other, while a small intra-cluster variance implies the points of each cluster are located close to their respective centroid. Consequently, a large optimal ratio is associated with a more clearly robust regime structure.

One of the long-standing issues with applying clustering methodologies to atmospheric data is the inability to choose a cluster number \emph{a priori}. A commonly used method for identifying cluster number \emph{a posteriori} is the Bayesian information criterion (BIC), which we will employ in this paper. This aims to minimise the residual unexplained variance of the dataset while attempting to account for over-fitting by penalising large numbers of free parameters \cite{Fraley1998}. In the context of K-means clustering, this takes the form:

\begin{equation}
    BIC=\sum_n^{N}\argmin_K \left( \left\lVert X_K-x_n\right\rVert^{2} \right)  +KD\cdot\log{(N)}
\end{equation}

The first term is the sum of squared distances of each of the $N$ datapoints to the nearest of the $K$ clusters, with centroids $X_K$ (the intra-cluster variance above). The second term is our parameter penalisation, where $D$ is the state-space dimension, and where the logarithmic scaling with $N$ follows from information theoretic arguments \cite{Schwarz1978}.

\section{Results}
\label{sec:results}

We now show how removing the influence of the jet speed alters the structure of the Z500 phase space and diagnose regime structure in this residual space.

\subsection{Removing the influence of the jet speed}
\label{sec:removing_speed}

The daily jet speed index enjoys statistically significant linear correlation with the first two PCs of Z500, at approximately $0.44$ and $0.49$ respectively. Correlations with further PCs decrease rapidly, implying that most of the jet speed variability is captured by the NAO (first PC) and the East Atlantic pattern (second PC), in line with the results of \citeA{Woollings2010b}: see Figure A1 in the Supporting Information (SI), included here as Appendix A.

Because the influence of the jet speed on the PCs is linear to good approximation, we remove it from the Z500 phase space using linear regression. Concretely, the jet speed is regressed against each of the 10 PCs separately; the residuals of this regression form the coordinate vectors of a new phase space where the jet speed, by construction, does not correlate with any of them. We call this the \emph{residual phase space}. 

\begin{figure}
     \noindent\includegraphics[width=\textwidth, scale=1.5]{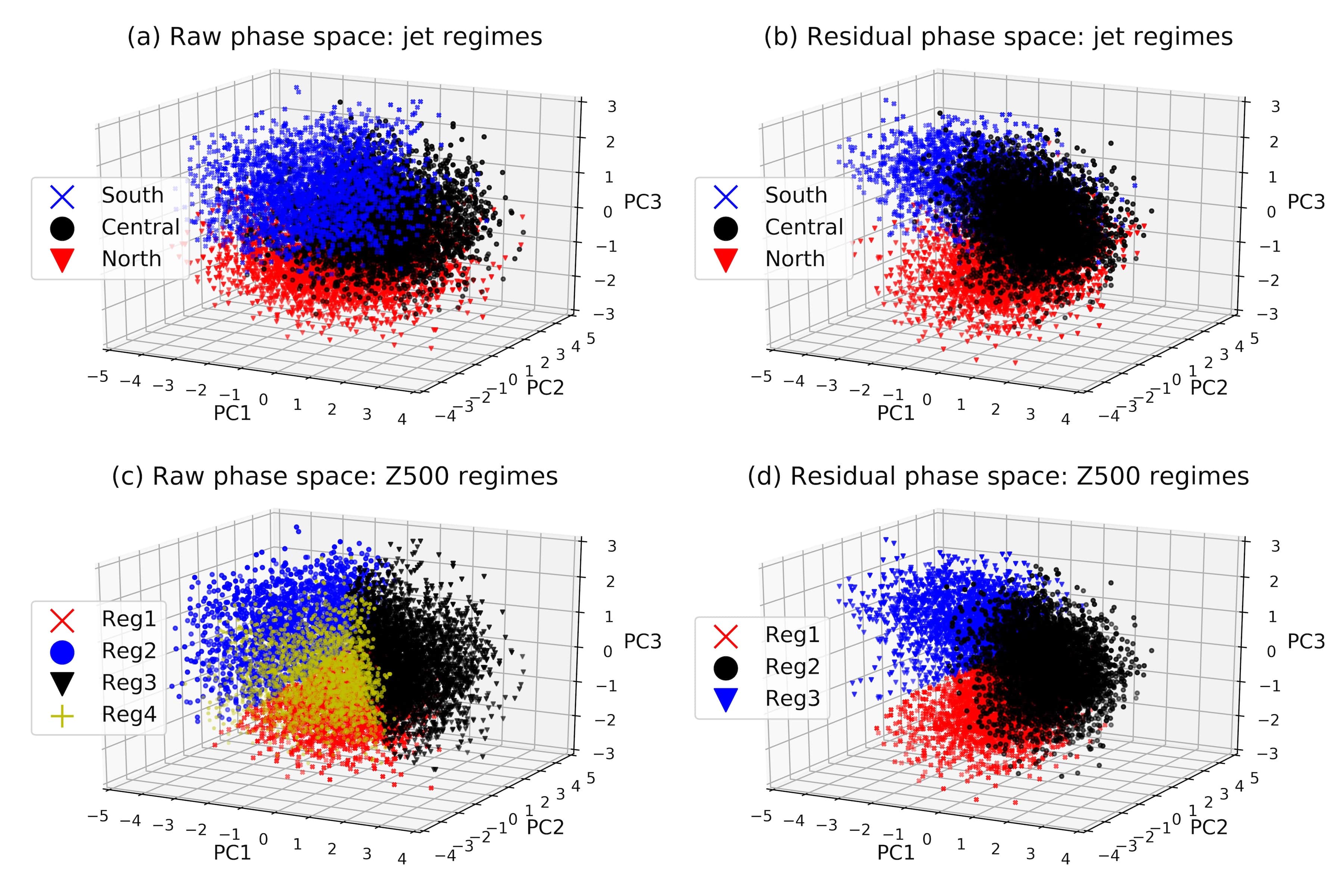}
    \caption{Projections of the Z500 phase space of ERA20C onto the first 3 EOFs. In (a) and (c) is shown the raw phase space with points marked according to the jet latitude regimes in (a) and according to the K-means cluster regimes in (c). In (b) and (d) is shown the residual phase space (i.e. with the jet speed removed), with points in (b) marked by jet latitude regimes and (d) by clustered regimes. In (c), the regimes shown are NAO+ (Reg1), NAO- (Reg2), Atlantic Ridge (Reg3) and Scandinavian Blocking (Reg4).}
    \label{fig:phase_space}
\end{figure}

The impact of this procedure can be seen in Figure \ref{fig:phase_space}. In (a) and (c) are shown a projection of the 10-dimensional phase space onto the first 3 EOFs: each point therefore corresponds to a particular Z500 flow pattern. In (a), the points have been coloured according to which of the three jet latitude regimes they belong to, while in (c), the colouring is done according to the regimes identified by K-means clustering with K=4. The phase space does not exhibit deviations from Gaussianity easily visible to the human eye. By contrast, (b) and (d) show the residual phase space, where non-Gaussian structure is clearly identifiable. Once more, in (b) the colour indicates the three jet regimes, while in (d), the colour indicates the regimes identified by K-means clustering of the residual phase space with K=3. It can be seen that the K-means clustering essentially reproduces the three jet regimes, a correspondence we will quantify in Section \ref{sec:regime_structure}. In fact, the curvature visible in Figure \ref{fig:phase_space}(b) and (d) corresponds exactly to the non-linear relationship between the jet latitude and the NAO first observed in \citeA{Woollings2010b}: see also Figure 2 of \citeA{Strommen2020}, where this non-linearity is more clearly highlighted. Note that a complementary version of Figure \ref{fig:phase_space} with no colouring of points can be found in the Appendix.

This lends compelling evidence towards the idea that the regimes in the Z500 phase space are simply an imprint of the three jet latitude regimes. However, it is still possible that K=3 is in some sense not the optimal choice of clusters for the residual phase space. We address this in the next section.

\subsection{Identifying the optimal regime structure}\label{sec:regime_structure}

\begin{figure}
     \noindent\includegraphics[width=\textwidth]{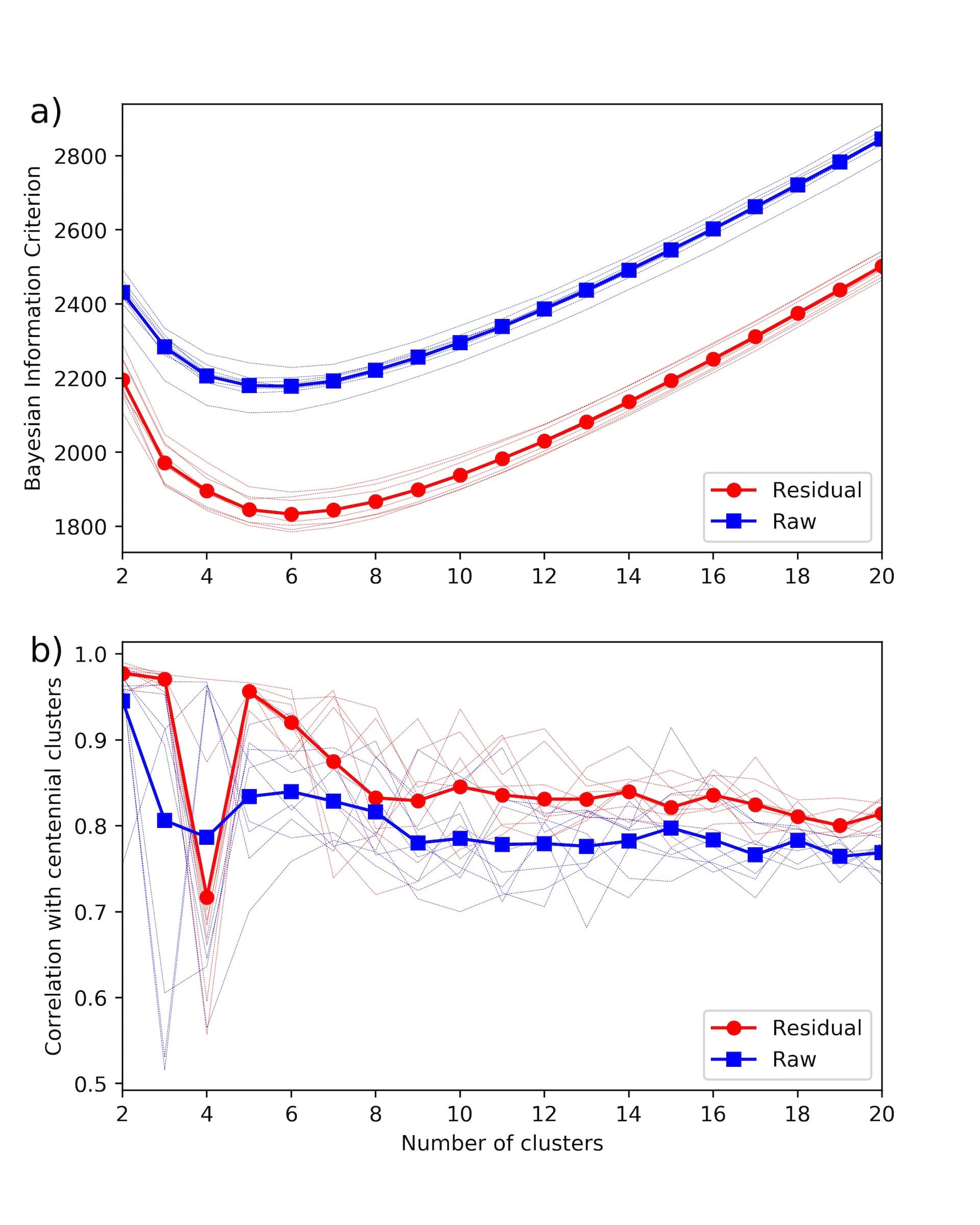}
    \caption{Top: The BIC for varying numbers of K-means clusters found in the residual phase space (red) and the raw phase space (blue). Thin lines show results for each of the 30-year windowed periods, with the mean indicated by the thick lines. Bottom: Thin lines show the average full-field pattern correlation of cluster composites between each 30 year window of data and the full centennial dataset. Again the thick lines indicates the mean of the windowed results.}
    \label{fig:choosing_K}
\end{figure}

We now consider the question of identifying an optimal number of regimes to use in this residual phase space. Due to the visible non-Gaussian structure, conventional significance testing based on the cluster sharpness \cite{Strommen2019a} is of no use here; significance is always 100\% against a null hypothesis of linear auto-correlated noise. Instead we consider two metrics for identifying the regime number. The BIC is an information theoretic measure of model suitability, as discussed in Section \ref{sec:data_and_methods}, and the minima of the BIC can be used to identify the optimal number of free parameters in a model, which in our case is precisely the cluster number. Figure \ref{fig:choosing_K}(a) shows that a broad minima exists for both raw and residual PCs when using K=6, which is consistent with the findings of, for example, \citeA{Falkena2019}. Both datasets have had their variance normalised prior to clustering and so the results can be directly compared. We see that in all cases, clusters found in the residual PC have a lower BIC than those in the raw PC space space, indicating a much better model fit to the data; the assumption of underlying multimodality is better justified. 

However, the BIC is only one of many proposed methodologies for selecting regime number which can often give conflicting results, and we must bear in mind its underlying formal assumptions that our data violates, particularly the conditions of stationarity and independence. We are therefore prompted to consider a more physically grounded, domain-specific metric that reflects the intended reason for identifying regimes in the first place. In our case, we would expect dynamically relevant regimes to be approximately stationary features of the mid-latitude circulation over centennial timescales, at least in terms of spatial patterns if not in residence times or transition probabilities. 

To that end, we define a stationarity metric by calculating the average area-weighted pattern correlation between clusters found in rolling 30-year windows of our dataset, and those found in the full centennial dataset. We use 9 windows, [1902-1931,1912-1941,...1982-2010], and in each case find the bijection between windowed and centennial clusters that maximises the pattern correlation of the regime composites.
The results of this are shown in Figure \ref{fig:choosing_K}(b), for both raw and residual PCs. 

Again we see a marked improvement when clustering in the residual space, with the stationarity almost always higher. Of particular note is that while for K$>$2 the raw PCs show no clear link between stationarity and regime number, the residual PCs show a clear distinction between very stationary clusters for K=2,3, and 5, a general decline for K$>$5, and a strong non-stationarity for K=4 (which is addressed in Subsection \ref{sec:K=5}).

This increased stability for almost all cluster numbers is in accordance with the observation in \citeA{Woollings2014} that while jet speed shows statistically significant inter-decadal variability, the jet latitude does not. Removing the influence of the jet speed would therefore be expected to reduce decadal variability.

\subsection{The Case K=3: Jet-analogue regimes}
\label{sec:K=3}

In Figure \ref{fig:K3_patterns} we look at the case K=3 in more detail. While it does not represent a minima of the Bayesian information criterion, the regime clusters are highly stable across the 20th century, and appear to correspond closely to the three jet latitude regimes. We therefore examine these clusters in more detail.

From the geopotential anomaly composites associated with each cluster, seen in Figure \ref{fig:K3_patterns}(a), we label these as a blocking/NAO+ hybrid (BLK/NAO+), a pure NAO- pattern, and an Atlantic ridge/NAO+ hybrid (AR/NAO+) by analogy to the 4 traditional Euro-Atlantic regime. This identification holds quantitatively, as we can see if we look at the coincidence matrix which encodes how many days assigned to a regime $i$ under one clustering framework are assigned to a regime $j$ under a different set of regime classifications:

\[
\label{mat:sim_U4_R3}
\bbordermatrix{
~&BLK&NAO+&NAO-&AR\cr
BLK/NAO+&2103&1785&65&229\cr
NAO-&264&128&1772&125\cr
AR/NAO+&102&1045&200&1992\cr 
}
\]

The corresponding zonal wind composites, displayed in Figure \ref{fig:K3_patterns}(b), show that the projection onto the jet is significant. Our NAO- cluster is linked to a Southerly jet, the Atlantic ridge  with a Northerly jet, while the blocking pattern is characterised mainly by a weak jet over Western Europe. 

The similarity matrix below shows how similar the assignment of days is for the jet latitude clusters and the clusters in residual Z500 obtained here. We see that approximately 60\% of the days assigned to each Z500 regime come from a single jet regime, as expected from the visual similarity apparent in Figure \ref{fig:phase_space}:
\[
\label{mat:sim_jli_R3}
\bbordermatrix{
~&Central&Southern&Northern\cr
BLK/NAO+&2290&606&1286\cr
NAO-&672&1483&134\cr
AR/NAO+&1209&141&1989\cr
}
\]

The statistics of regime lifetime, displayed in Figure \ref{fig:K3_patterns}(c), show a strongly Markovian persistence structure, with a typical lifetime of around 1 week, and with rare month-long regime events occurring for all clusters at some point in the centennial record. Given the strong Markovianity, there is value in examining the transition matrix, which defines the daily probability of moving from one cluster to another:

\[
\bordermatrix{
  From\downarrow To\rightarrow &BLK/NAO+&NAO-&AR/NAO+\cr 
  BLK/NAO+&0.85&0.04&0.11 \cr
  NAO-&0.08&0.86&0.06 \cr
  AR/NAO+&0.13&0.05&0.82\cr
}
\]

This is strongly inhomogeneous (a desirable trait if regimes are to be used for predictive purposes), with transitions into the NAO- state much less common than transitions between the two hybridised NAO+ states. This bears similarities to the transition matrix found for clusters of the jet latitude index such as in \cite{Franzke2011}, where the Northern and Central jet states prefer to transition to each other. 

\subsection{The Case K=5: Extended jet regimes}
\label{sec:K=5}

The variability of the jet latitude seems to be well described by a trimodal structure, but if we also want to capture the range of Atlantic blocking dynamics, we may benefit from extending our basis of regimes. After all, Scandinavian blocking shows trademarks of a quasi-persistent regime state, and persistent blocking in general provided the motivation behind early work on weather regimes \cite{Vautard1990}.

Strong stationarity also appeared in Figure \ref{fig:choosing_K} for K=5, which also has the advantage of being a near minima of the BIC, and so we examine the resulting regime composites in Figure \ref{fig:K5_patterns}. 

Here we see that clusters 1, 4 and 5 map clearly onto the Central, Northern and Southern jet latitude regimes respectively, while clusters 2 and 3 represent flows with a weak jet over Europe, and European blocking characteristics. 
The two new regimes, clusters 1 and 2 here, are hybrids: cluster 1 is 64\% AR, 26\% BLK/NAO+ and 10\% NAO-, while cluster 2 is 57\% NAO+/BLK and 39\% NAO-. We call these a Norwegian low and high respectively. This information can be read off from the similarity matrix:

\[
\bbordermatrix{
  ~&NOR-&NOR+&BLK&NAO-&AR\cr 
  BLK/NAO+&570 & 1038 & 3509 & 0 & 65 \cr
  NAO-&223 & 709 & 0 & 1339 & 18 \cr
  AR/NAO+&1397 & 81 & 42 & 49 & 1770 \cr
}
\]

Hence, we see that the 5 regimes are a straightforward extension of the 3 regimes in the previous section, adding extra detail but not removing any prior patterns. The full transition matrix is shown in Table A1. The ability to capture both the full set of jet regimes and blocking patterns in a stable and reproducible way is a noteworthy feature that derives from our residual clustering approach, and sets it apart from previous studies on regime identification. A visualisation of how the 5 clusters are positioned in phase space is provided in Figure A3 of the SI.

When 4 clusters are requested, K-means clustering consistently returns clusters 3, 4 and 5 of Figure \ref{fig:K5_patterns}, but in different 30 year windows switches between including clusters 1 and 2 (not shown). The resulting inconsistency is the cause of the very high non-stationarity visible in Figure \ref{fig:choosing_K}. This can be seen as additional support to the view that we are detecting significant physically consistent states; the clustering algorithm is by no means indifferent to the number of clusters we specify.

\section{Conclusions}
\label{sec:conclusions} 

In this work we have considered the hypothesis that the near-Gaussian variability of the jet speed is a confounding factor that obscures the non-linear regime structure of the Euro-Atlantic circulation. Once the jet speed has been regressed out of the principal components of Z500, a clear, visual non-linearity is instantly apparent in phase space. We found that K-means clustering applied in the standard way to this residual Z500 demonstrated marked improvements in statistical significance, in the Bayesian Information Criterion, in identifying regime number, and perhaps most importantly in inter-decadal stability across the 20th Century. 

It was shown that three regimes managed to capture a Northern and Southern jet as well as a weakly blocked state, significantly overlapping the previously identified three jet latitude regimes. Thus we have managed to provide a framework which straight forwardly unifies both jet and circulation regimes. This framework adds considerable clarity to earlier approaches at such a unification \cite{Madonna2017}. Furthermore by extending to five regimes, we find more strongly blocked states linked to a tilted and split jet in addition to the three traditional jet regimes, which also show strong stationarity.

\begin{figure}
     \noindent\includegraphics[width=\textwidth]{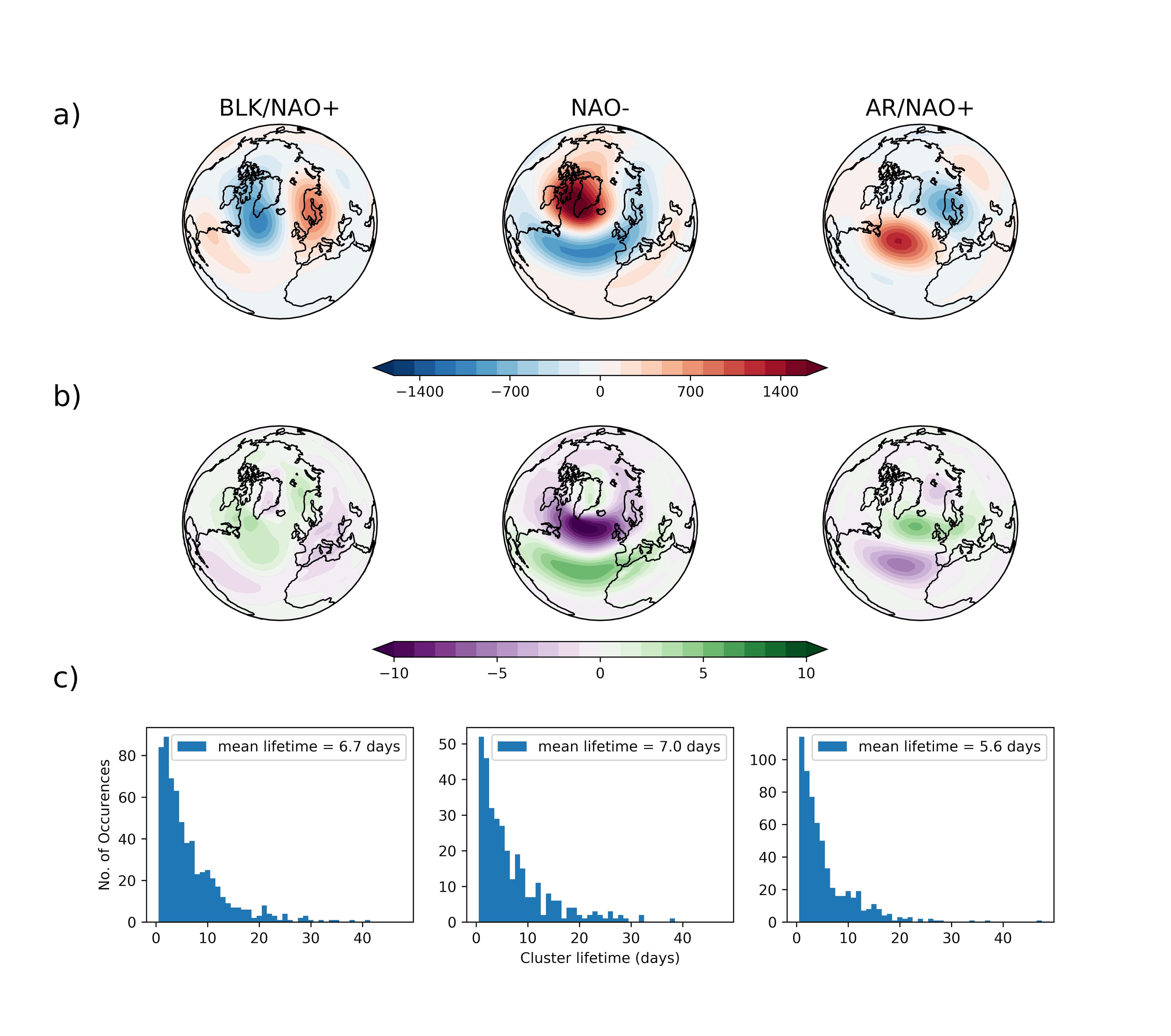}
    \caption{Clusters composites for K-means clusters found using K=3 in residual PCs of DJF Z500 1902-2010. Composites of Z500 anomalies for days assigned to each regime are shown in \textbf{a)}, while composites of U850 anomalies are shown in \textbf{b)}. Histograms of cluster lifetime are shown in \textbf{c)}.}
    \label{fig:K3_patterns}
\end{figure}

\begin{figure}
     \noindent\includegraphics[width=\textwidth]{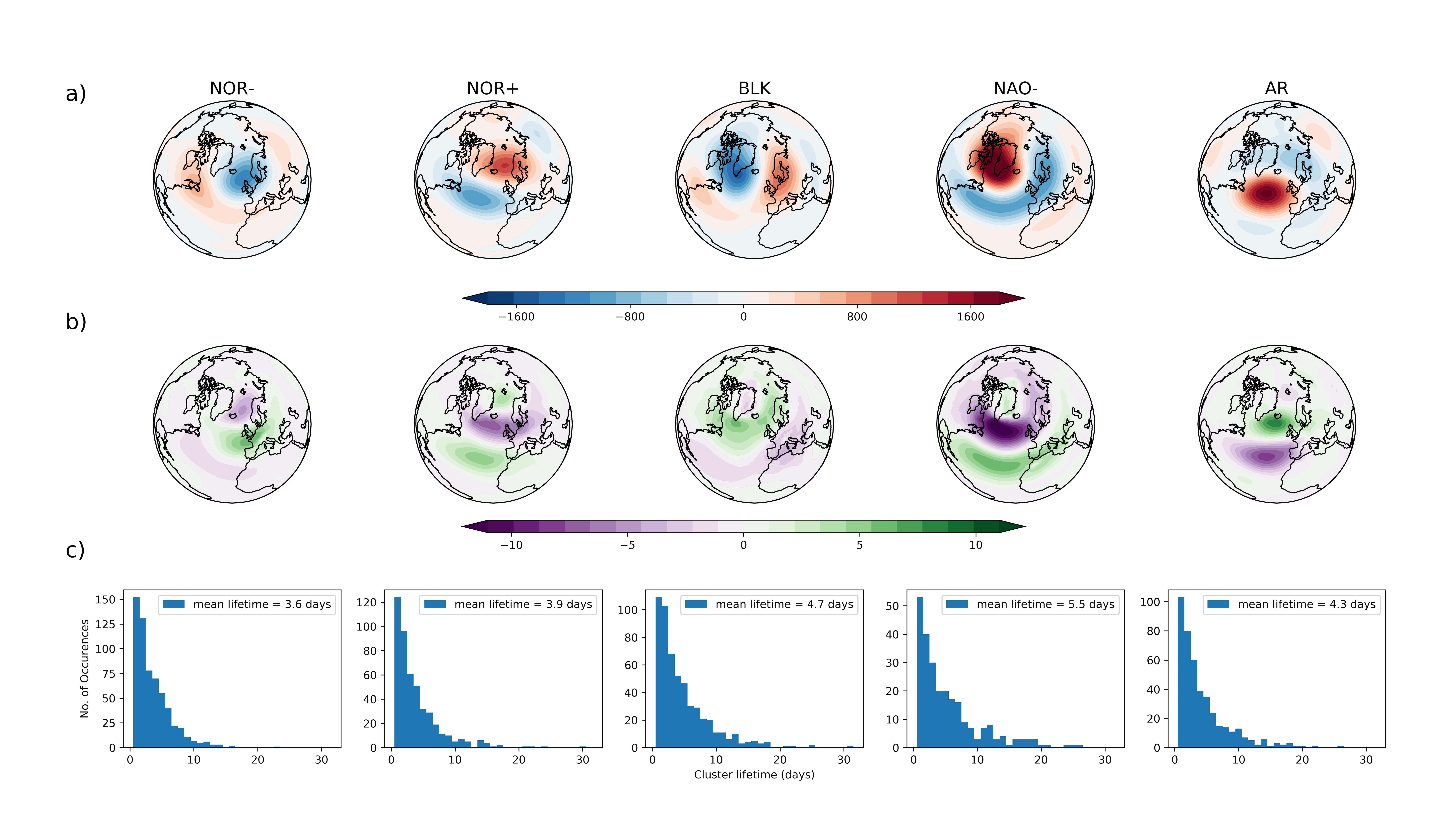}
    \caption{As for Figure \ref{fig:K3_patterns} but using K=5.}
    \label{fig:K5_patterns}
\end{figure}

The issue of unambiguous regime identification, and of strong sensitivity of regime patterns to data and parameter changes, has caused considerable consternation and scepticism over the usefulness of the regime framework in the past. We have shown that many of these issues can essentially be removed entirely by removing the jet speed. We note that this connection between zonal wind speeds and regime structure has precedent in the literature. In \citeA{Woollings2008}, a bimodal framework of regimes was presented based on Rossby wave-breaking. The positive NAO, corresponding to strong zonal flow, was envisaged as being the `generic', undisturbed state, while the negative NAO corresponded to periods of frequent blocking events. In other words, strong zonal flow (NAO+) is in some sense not a regime on its own, and once this is removed you recover a range of blocking patterns. This is entirely consistent with the clusters obtained in both Figure \ref{fig:K3_patterns} and \ref{fig:K5_patterns}, which omit a strong NAO+ pattern in favour of various blocking patterns. The dependence of regime structure on wind speeds is also explicit in \citeA{Benzi1986} and \citeA{Ruti2006}, where evidence of bimodality only emerges for specific ranges of zonal wind forcing. Our work therefore corroborates the idea that regime structure is best understood in the absence of wind speed forcing.

With this increased regime reproducibility comes increased confidence in the ability to use regime statistics as model diagnostic tools as well as potentially providing insight into the underlying skeletal structure of non-linearity over the  Euro-Atlantic sector. In particular, this new clustering framework might be particularly suited for identifying the action of teleconnections and external forcing patterns on the North Atlantic, which can be difficult to study with either non-stationary datasets or data confined to a short, modern period.

%
%

\acknowledgments
KJS was funded by a Thomas Philips and Jocelyn Keene Junior Research Fellowship in Climate Science at Jesus College, Oxford. JD is funded by NERC Grant NE/L002612/1. The ERA20C data used can be accessed at \url{https://www.ecmwf.int/en/forecasts/datasets/reanalysis-datasets/era-20c}, courtesy of the European Center for Medium-Range Weather Forecasts.



%
%
\newpage
\bibliography{references}

%
%
\newpage

\clearpage
\appendix
\section{Supporting Information}

The Appendix contains the Supporting Information associated with this article, namely Figures A1 through A3 and Table A1.

 \begin{figure}
     \noindent\includegraphics[width=\textwidth, scale=1.5]{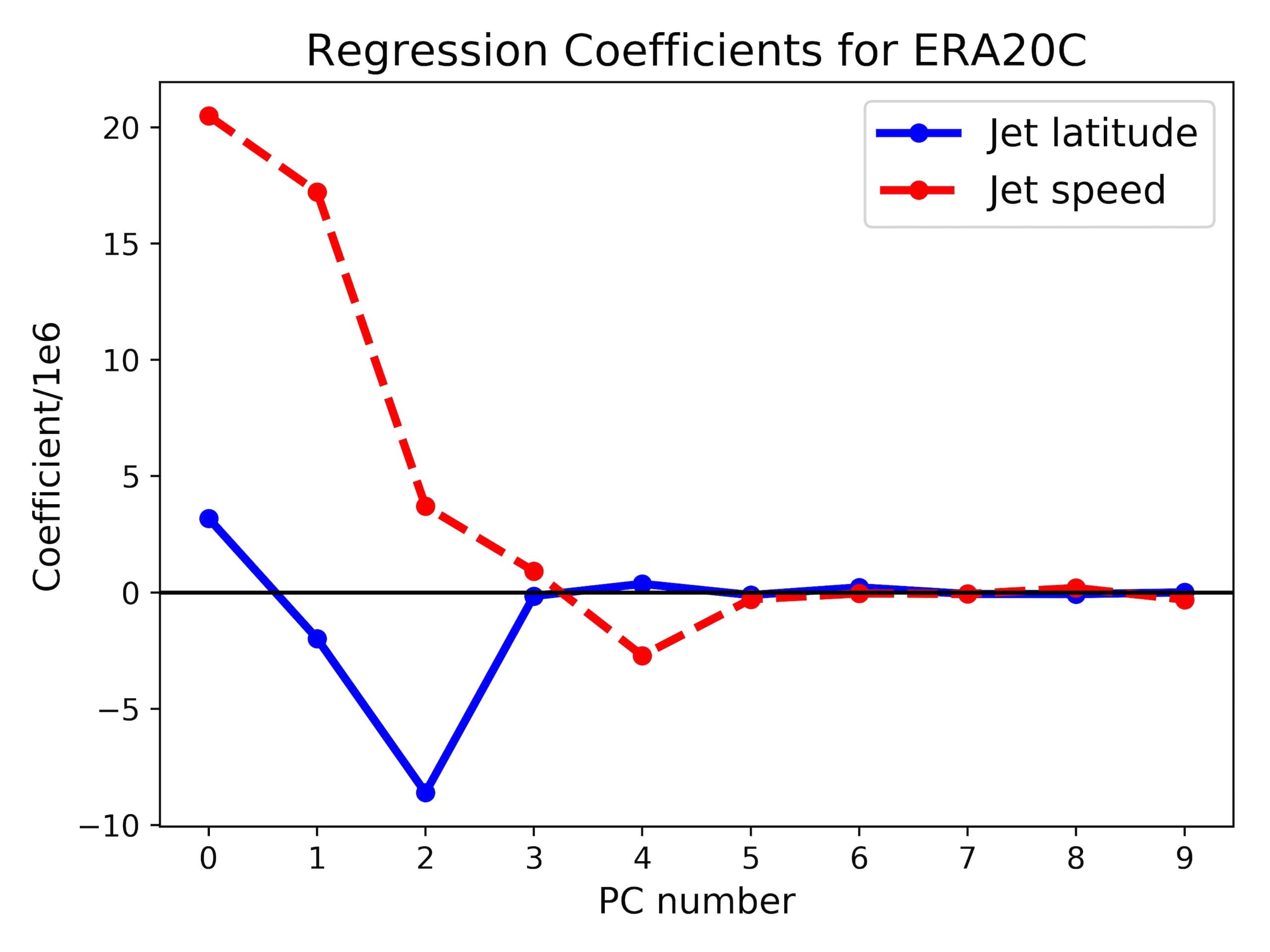}
    \caption{Regression coefficients between the jet speed (red, dashed), jet latitude (blue, solid) and the 10 leading principal components of detrended Z500 anomalies in the Euro-Atlantic region.}
    \label{fig:regr_coefs}
\end{figure}

 \begin{figure}
     \noindent\includegraphics[width=\textwidth, scale=1.5]{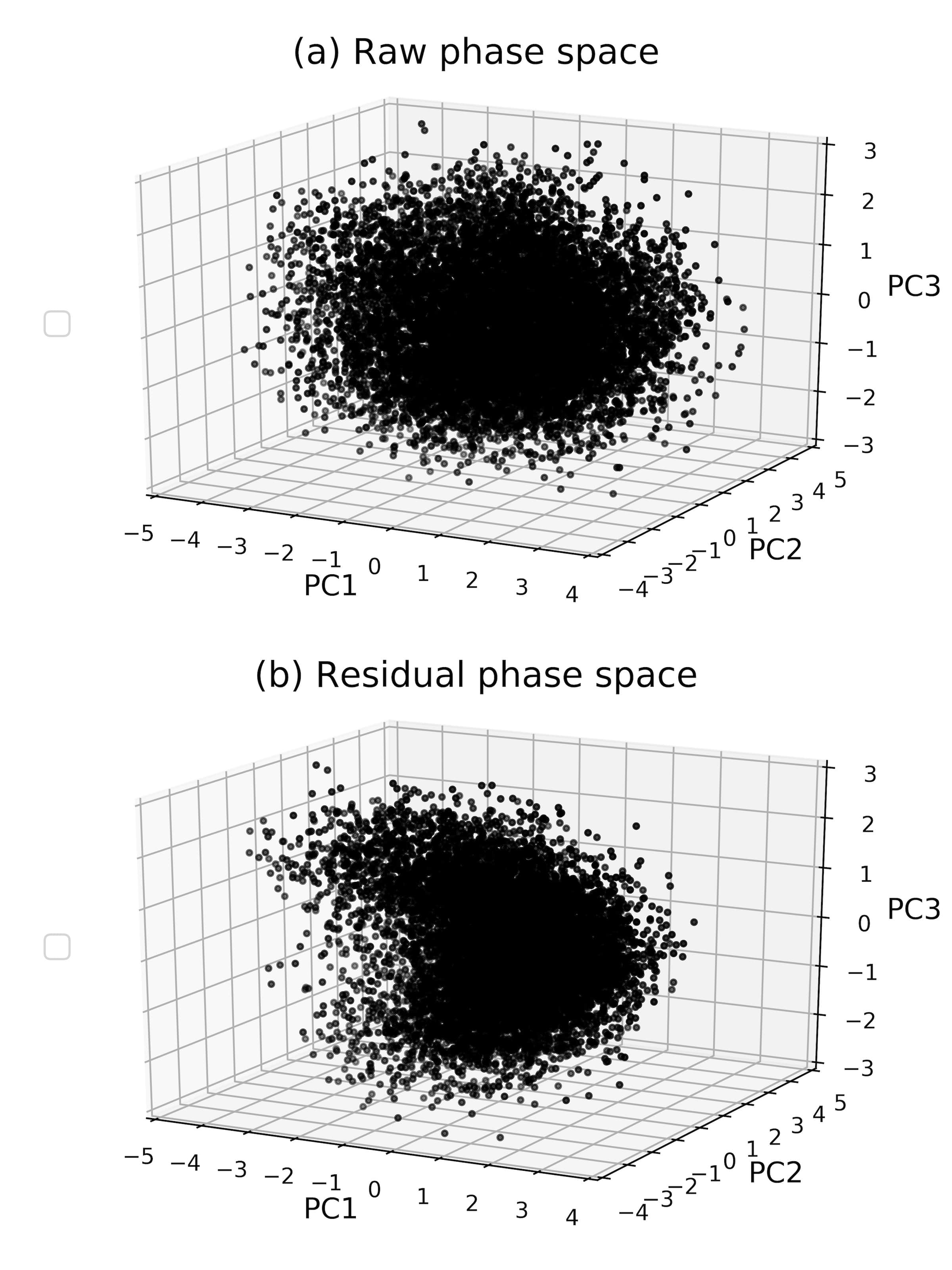}
    \caption{Projections of the Z500 phase space of ERA20C onto the first 3 EOFs. In (a) for the raw PCs and (b) for the residual PCs (i.e. with the jet speed removed).}
    \label{fig:raw_phase_space}
\end{figure}

 \begin{figure}
     \noindent\includegraphics[width=1.0\textwidth, scale=1.0, height=0.9\textheight]{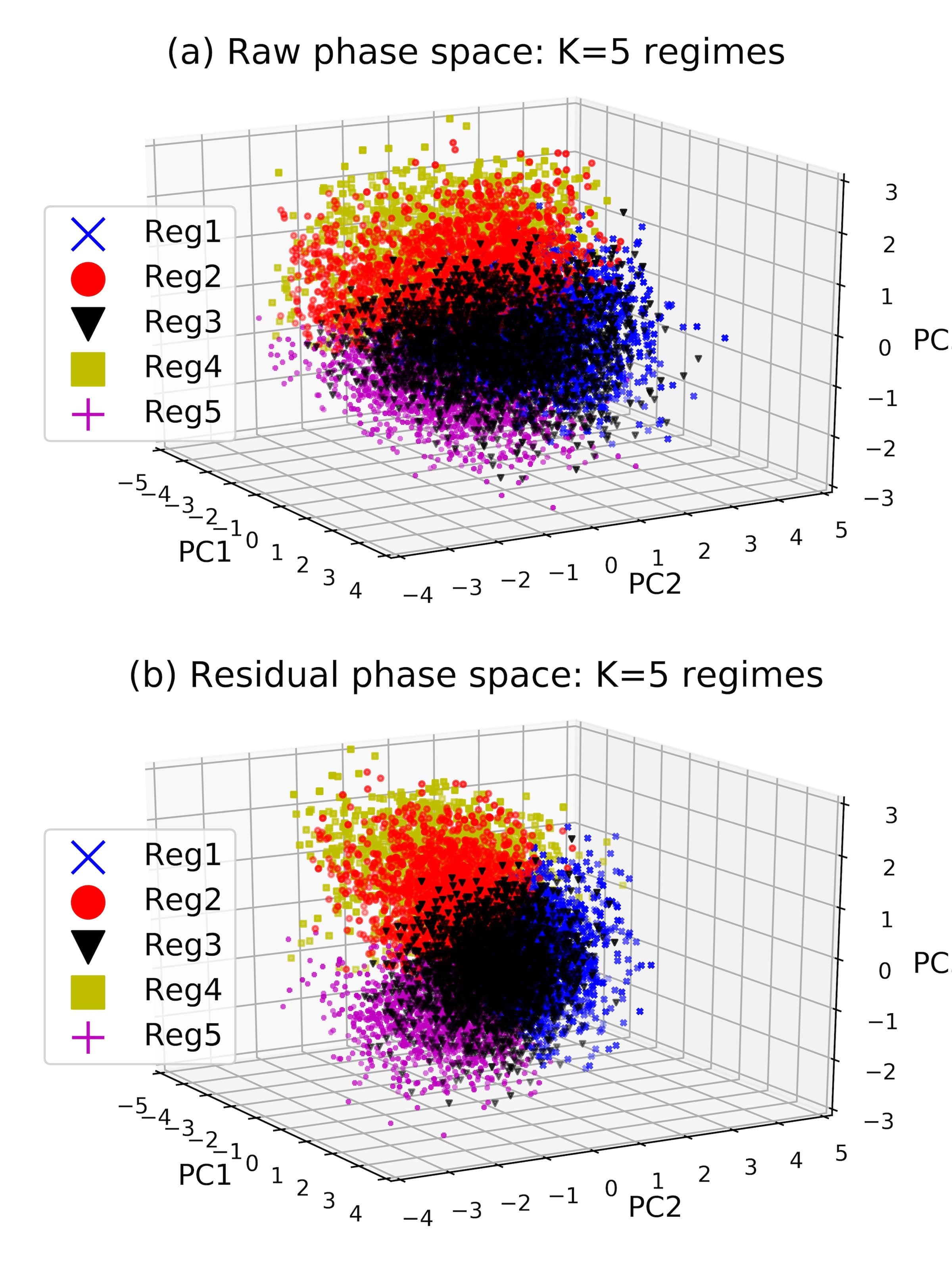}
    \caption{Projections of the Z500 phase space of ERA20C onto the first 3 EOFs. In (a) for the raw PCs and (b) for the residual PCs (i.e. with the jet speed removed). In both cases, points are marked according to which of the K=5 regimes they belong to. Using the naming from the main document, the clusters are: NOR- (Reg1), NOR+ (Reg2), BLK (Reg3), NAO- (Reg4) and AR (Reg5).}
    \label{fig:K5_phase_space}
\end{figure}

\begin{table}[]
    \centering
    \begin{tabular}{|c|c|c|c|c|c|}
    \hline
    From$\downarrow$ To$\rightarrow$&NOR-&NOR+&BLK&NAO-&AR\\
    \hline
    NOR-&0.72&0.07&0.08&\textcolor{red}{0.05}&0.08\\ 
    \hline
    NOR+&0.05&0.74&\textbf{0.11}&0.06&\textcolor{red}{0.03}\\ 
    \hline
    BLK &0.08&0.06&0.79&\textcolor{red}{0.00}&0.06\\ 
    \hline
    NAO-&0.07&0.08&\textcolor{red}{0.00}&0.82&\textcolor{red}{0.03}\\ 
    \hline
    AR  &\textbf{0.10}&\textcolor{red}{0.02}&0.08&\textcolor{red}{0.02}&0.77\\ 
    \hline
    \end{tabular}
    \caption{The matrix of daily transition probabilities between the 5 regimes shown in figure 4 of the main article. Preferred transitions are marked in bold, while avoided transitions are in red.}
    \label{tab:K5_transmat}
\end{table}

\end{document}